\documentclass[twocolumn,showpacs,preprintnumbers,amsmath,amssymb]{revtex4}

\begin{document}

\draft

\title{ 
On the applicability of an equation of motion method  
at low-temperatures: \\
comments on cond-mat/0309458 and cond-mat/0308413 
}

\author{Akira Oguri}
\address{
          Department of Material Science,
          Osaka City University, 
          Sumiyoshi-ku, Osaka 558-8585,
          Japan 
         }

\date{\today}


\begin{abstract}
The equation of motion method (EOM) 
is one of the approximations to calculate transport coefficients of 
interacting electron systems. 
The method is known to be useful to examine 
high-temperature properties. 
However, sometimes a naive application of the EOM fails 
to capture an important physics at low-energy scale, and
it happens in recent preprints cond-mat/0309458 and cond-mat/0308413
which study a series of quantum dots.
These preprints concluded that 
a unitarity-limit transport due to the Kondo resonance, 
which has been deduced from a Fermi-liquid 
behavior of the self-energy at $T=0$, $\omega=0$ 
[A.O., PRB {\bf 63}, 115305 (2001)], does not occur. 
We show that the EOM self-energy obtained with a finite cluster  
has accidentally a singular $1/\omega$ dependence 
around the Fermi energy, and it misleads one to 
the result incompatible with a Fermi-liquid ground state.

\end{abstract}

\pacs{72.10.-d, 72.10.Bg, 73.40.-c}

\maketitle
 
\narrowtext

\begin{figure}[b]
\begin{center}
\setlength{\unitlength}{0.65mm}

\begin{picture}(130,20)
\thicklines
\put(5,10){\line(1,0){3}}
\put(12,10){\line(1,0){6}}
\put(22,10){\line(1,0){6}}
\put(32,10){\line(1,0){6}}

\multiput(42.5,10)(2,0){3}{\line(1,0){1}}
\multiput(82.5,10)(2,0){3}{\line(1,0){1}}

\put(50,10){\line(1,0){30}}
\put(92,10){\line(1,0){6}}
\put(102,10){\line(1,0){6}}
\put(112,10){\line(1,0){6}}
\put(122,10){\line(1,0){3}}

\put(10,10){\circle{4}} 
\put(20,10){\circle{4}} 
\put(30,10){\circle{4}} 
\put(40,10){\circle{4}} 
\put(50,10){\circle*{4}} 
\put(60,10){\circle*{4}} 
\put(70,10){\circle*{4}} 
\put(80,10){\circle*{4}} 
\put(90,10){\circle{4}} 
\put(100,10){\circle{4}} 
\put(110,10){\circle{4}} 
\put(120,10){\circle{4}}

\put(54,16){\makebox(0,0)[bl]{\Large $t$}}
\put(64,16){\makebox(0,0)[bl]{\Large $t$}}
\put(74,16){\makebox(0,0)[bl]{\Large $t$}}
\put(42.5,15){\makebox(0,0)[bl]
{\Large $v_L^{\phantom{\dagger}}$}}
\put(82,15){\makebox(0,0)[bl]
{\Large $v_R^{\phantom{\dagger}}$}}

\put(39,0){\makebox(0,0)[bl]{\Large $0$}}
\put(49,0){\makebox(0,0)[bl]{\Large $1$}}
\put(59,0){\makebox(0,0)[bl]{\Large $2$}}
\put(66.5,0){\makebox(0,0)[bl]{\Large $\cdots$}}
\put(77.5,0){\makebox(0,0)[bl]{\Large $N$}}

\end{picture}

\end{center}

\caption{Schematic picture of the model: 
($\bullet$) Hubbard chain of $N$ sites, and 
($\circ$) noninteracting leads.  The tunneling 
matrix elements $v_L$ ($v_R$) connects the cain 
and the left (right) lead to make the energy 
spectrum of the whole system continuous.} 
\label{fig:model}
\end{figure}
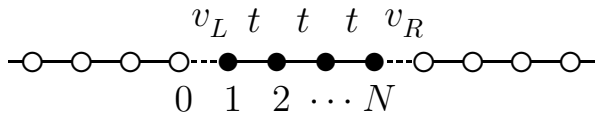

\section{Background}
\label{sec:background}

A Hubbard chain of finite size $N$ connected to two noninteracting 
leads, which is illustrated in Fig.\ \ref{fig:model}, 
has been studying 
as a model for a series of quantum dots \cite{chiappe,busser,ao7,ao9,ao10}.
Here $N$ corresponds to the number of quantum dots and should be small, 
while the number of the sites in the noninteracting leads at 
left and right are infinite, t.e., $N_L\to \infty$ and $N_R\to \infty$.  
It has been shown for the chain of an odd $N$  
that the transmission probability 
reaches a unitarity-limit value at $T=0$ if 
the system has the electron-hole symmetry (half-filling)  
together with the inversion 
symmetry ($v_L^{\phantom{0}}$ = $v_R^{\phantom{0}}$)  \cite{ao9}.
This conclusion has been deduced from a Fermi-liquid behavior 
of the self-energy summarized simply in Eq.\ (\ref{eq:Fermi_ground}). 

Recently, the dc conductance has 
been calculated with an extended version of 
the equation of motion method (EOM) \cite{MWL}, 
which is essentially the Hubbard I approximation \cite{hubbardI}, 
at zero temperature \cite{chiappe,busser}. 
The results of these preprints contradict with 
the unitarity-limit transport mentioned above.
The purpose of this report is to show that
the discrepancy is caused just by a naive use 
of the EOM self-energy of obtained with a finite cluster,
which accidentally has a singularity at low-energies and 
is not compatible with the Fermi-liquid ground state.

In Sec.\ \ref{sec:conductance}, 
we describes briefly the results 
deduced from a Fermi-liquid property 
Eq.\ (\ref{eq:Fermi_ground}) to make our points clear.
Then, in Sec.\ \ref{sec:EOM}, 
the properties of the approximate self-energy 
in the EOM are discussed.

\section{conductance at zero temperature}
\label{sec:conductance}

\subsection{Formulation}

The dc conductance at $T=0$ is determined by 
the value of a Green's function $G_{N 1}^{+}(\omega)$
at the Fermi energy $\omega=0$ \cite{ao9};
\begin{equation}
   g_{N}^{\phantom{\dagger}} \,=\, {2 e^2 \over h} \ 
    4\, 
        \Gamma_R(0) \,\Gamma_L(0) 
       \, \left|\, G_{N 1}^{+}(0)\,\right|^2  
               \;, 
\label{eq:cond} 
\end{equation}
Here 
$G_{ij}^{+}(\omega)$ is the retarded Green's in the
real space ($1\leq i j \leq N$), see Fig.\ \ref{fig:model}, 
and  $\Gamma_{\alpha}(\omega) = -\, 
v_{\alpha}^2 \mbox{Im} 
\, \mbox{\sl g}_{\alpha}^{+}(\omega)$
for $\alpha = L,\, R$ with 
$\mbox{\sl g}_L^+$ ($\mbox{\sl g}_R^+$) 
being an interface Green's function 
for the semi-infinite lead at left (right).  
The Dyson equation can be written in a $N \times N$ matrix form; 
\begin{eqnarray}
\left\{\mbox{\boldmath $G$}(\omega)\right\}^{-1}  
  &=&  
\{ \mbox{\boldmath $G$}^{0}(\omega) \}^{-1}
- \mbox{\boldmath $\Sigma$}(\omega)   \;.
\label{eq:Dyson}
\end{eqnarray}
Here $\mbox{\boldmath $G$}^{0} = \{G^{0+}_{ij}\}$ 
is the noninteracting Green's function for $U=0$ defined with 
respect to the total system connected via tunneling  $v_L$ and $v_R$.
The inverse matrix of the unperturbed Green's function can be 
expressed as 
\begin{eqnarray}
\{ \mbox{\boldmath $G$}^{0}(\omega) \}^{-1} &=&
\omega \, \mbox{\boldmath $1$} 
-  \mbox{\boldmath $H$}_C^0  
- \mbox{\boldmath $V$}_{\rm mix}(\omega) \;,   
\\
\nonumber \\
 \mbox{\boldmath $H$}_C^0 \ &=& \  
\left [ \,
 \begin{matrix}
     0    & -t      &  & 0              \cr
            -t        & 0 &  \ddots       &               \cr
                    & \ddots              & \ddots  & -t  \cr 
            0                &               &  -t       & 0 \cr
    \end{matrix}        
          \, \right ]  
\;,
\\
\nonumber
\\
\mbox{\boldmath $V$}_{\rm mix}(\omega)   &=&  
 \left [ \, 
 \begin{matrix} 
 v_L^{2} \, \mbox{\sl g}_L^+(\omega) & 0 & \cdots &0 & 0 \cr
 0           &    0     & \cdots   &  0 & 0     \cr
 \vdots      &  \vdots  & \ddots   &  \vdots & \vdots  \cr 
 0 & 0 & \cdots & 0 & 0 \cr
 0 & 0 & \cdots & 0 & v_R^{2} \, \mbox{\sl g}_R^+(\omega) \cr
\end{matrix}                    
 \, \right ]  
\label{eq:V_mix}
\;.
\end{eqnarray}
Here $\mbox{\boldmath $1$}$ is the $N \times N$ unit matrix.
The self-energy due to the Coulomb 
interaction $\mbox{\boldmath $\Sigma$}$ is also defined with 
respect to the total system of continuous energy spectrum.
Note that the Green's function $G_{1N}^{+}(0)$ appearing 
in Eq.\ (\ref{eq:cond})
can be obtained by taking the inverse of 
the matrix in Eq.\ (\ref{eq:Dyson}).

\subsection{Fermi-liquid behavior}

At $T=0$, $\omega=0$ not only the Fermi-liquid 
but also some other \lq\lq liquid" states of interacting 
electron systems show a behavior, 
\begin{equation}
\mbox{Im}\, \mbox{\boldmath $\Sigma$} (0) =0 \;, 
\quad \mbox{and} \quad \  
\mbox{Re}\, \mbox{\boldmath $\Sigma$} (0) \ \,  \mbox{is \/ {\em finite}}. 
\label{eq:Fermi_ground}
\end{equation}
This property of the self-energy,  
the electron-hole symmetry, and the inversion symmetry 
are the three keys to lock the conductance for odd $N$ 
at the unitarity-limit value.
No other assumptions have been made in the proof \cite{ao9}.
The perturbation expansion in $U$ of Yamada-Yosida \cite{YamadaYosida} 
preserves the properties of Eq.\ (\ref{eq:Fermi_ground}) also 
for $N \geq 2$ owing to 
the presence of the noninteracting leads. 
The contributions of the 
hybridization $\Gamma_{L}$ and $\Gamma_R$  
in the unperturbed part $\mbox{\boldmath $G$}^{0}$ 
make it possible to take into account 
the virtual processes in which electrons  
visit inside the noninteracting leads of continuous spectrum 
during the multi-scattering by $U$.

For even $N$, the conductance decreases with increasing $N$ showing 
a tendency towards the development of a Mott-Hubbard 
insulator gap  $\Delta_{\rm gap}$  \cite{ao9,ao10}. 
Although we are interested in the system of small $N$,
a question arises: what can we see in the limit of $N\to \infty$?
Note that the results mentioned above hold just at $T=0$.
To describe the thermodynamic limit properly, 
we must consider the physics 
at finite but sufficiently low temperatures.
For odd $N$, 
the Kondo resonance of width $T_K$ is situated 
in the Hubbard gap. 
As $N$ increases, 
$T_K$ decreases and finally $T_K \to 0$ at $N \to \infty$ 
 because the developing gap disturbs 
the electrons to form the Kondo singlet. 
Thus, the value of the conductance $g_{N}^{\phantom{\dagger}}$ for 
odd $N$ ($=2M+1$) must depend on the order to take the two limit 
 $N \to \infty$ and $T \to 0$:
\begin{itemize}
\item
if  we take $T\to 0$ first and then $N\to \infty$,
 we see the physics at  $T<T_K$, i.e., the Kondo behavior, 
 \[
 \lim_{M\to\infty} \left[\,\lim_{T\to 0} g_{2M+1}^{\phantom{\dagger}} 
 \,\right]\,=\, 2e^2/h \;,
\]

\item
if  we take $N\to \infty$ first and then $T\to 0$, 
we see the physics at  $T >  T_K$, i.e., Mott-Hubbard behavior,
\[
 \lim_{T\to 0} \left[\,\lim_{M\to\infty} g_{2M+1}^{\phantom{\dagger}}
  \,\right] \, = \, 0 \;.
\]
\end{itemize}
Note that, for even $N$ ($=2M$), the result does not depend 
on the order to take these two limits,  
\[
\lim_{\scriptstyle M\to\infty \atop \scriptstyle T \to 0}
\, g_{2M}^{\phantom{\dagger}}
 \, = \, 0 \;.
 \]
At the temperatures $T_K < T < \Delta_{\rm gap}$,  
we see the insulating behavior for large $N$ independent 
of whether $N$ is even or odd. 
To observe the Kondo behavior at
an accessible temperature $T$, 
the resonance width $T_K$ should be sufficiently large,
so that the number of the quantum dots $N$ should be small.

\section{Applicability of EOM  at low temperatures}
\label{sec:EOM}

\subsection{Method}

In the EOM calculation of Refs.\ \cite{chiappe} and  \cite{busser}
the self-energy has been calculated 
using a finite cluster of the size $N_{\rm cl}$ decoupled 
from the total system.
Typically, the cluster introduced for the calculation 
consists of $N$ interacting sites,
several noninteracting sites in the left lead  $N_L'$, 
and that in the right lead $N_R'$; 
\begin{equation} 
N_{\rm cl} \ = \ N + N_{L}' + N_R' \;.
\end{equation} 
Using the ground state of the decoupled cluster 
obtained with an exact diagonalization, 
a $N_{\rm cl} \times N_{\rm cl}$ cluster  
Green's function $\widehat{g}_{\rm cl}$ 
has been calculated \cite{chiappe,busser}.
The EOM self-energy $\widehat{\Sigma}_{\rm cl}$ can be 
obtained from the Dyson equation for the decoupled cluster,
\begin{equation} 
\widehat{\Sigma}_{\rm cl}(\omega)
\ = \ \left\{\, \widehat{g}^0_{\rm cl}(\omega) \,\right\}^{-1} 
\,- \, 
\left\{\, \widehat{g}_{\rm cl}(\omega) \,\right\}^{-1}
\;.
\label{eq:DysonEOM}
\end{equation} 
Here $\widehat{g}^0_{\rm cl}$ is the $U=0$ Green's function for 
the decoupled cluster, which describes a discrete energy spectrum. 
Then, an approximate Green's function which takes into account 
the rest of the system can be calculated 
by substituting  $\widehat{\Sigma}_{\rm cl}(\omega)$ for 
the self-energy  $\mbox{\boldmath $\Sigma$}(\omega)$ defined 
with respect to the total system.

The applicability of this method is determined by the separation 
of the eigenvalues of the cluster $\Delta \epsilon$.
This is because $\,\widehat{\Sigma}_{\rm cl}$ has been calculated for 
the discrete energy spectrum, and 
the correlation effects on the low-lying states 
 below this energy separation have not been taken into account.
Therefore, the application of EOM can be justified 
at high-energies (or high-temperatures) 
above the energy separation $\Delta \epsilon$. 
The information 
about the many-body effects on the low-lying states 
are still missing  even though one gets a continuous spectrum 
after taking into account the connection to the rest of the system. 
Therefore, if a fine structure is seen in the obtained continuous spectrum, 
that structure is expected to be real only when the energy scale 
is consistent with the resolution $\Delta \epsilon$.
In principle, the energy resolution can be improved systematically   
by increasing the size of the cluster successively, 
and the numerical renormalization 
group (NRG) can be regarded as one of such approaches \cite{Wilson,KWW}. 
Since the cluster is introduced just for the approximation 
and have no physical meanings, 
the results of the observable quantities 
must not have a strong dependence on the value of $N_{L}'$ and $N_R'$.

\subsection{The $1/\omega$ singularity of the EOM self-energy}

In Refs.\ \cite{chiappe} and  \cite{busser}, 
the calculations have been carried out for 
the cluster with an odd number of sites $N_{\rm cl}$. 
At half-filling, the ground sate of the cluster 
is doublet with the total spin $S_{\rm tot}=1/2$. 
According to the appendix of Ref.\ \cite{busser},
the cluster Green's function  $\widehat{g}_{\rm cl}$  is defined 
using the equal-weight sum of 
$S_{\rm tot}^z = 1/2$ and $-1/2$ in Ref.\ \cite{busser},
while the work in Ref.\ \cite{chiappe} considers only the cluster 
state with $S_{\rm tot}^z = 1/2$.  
Furthermore, in Fig.\ 9 of Ref.\ \cite{busser}, 
these two procedures are shown to lead quite different results. 
This means that huge ambiguities exist 
in the EOM results at low energies.
In these two procedures, however, the equal-weight sum seems to be better
because it restores the rotational symmetry, 
so that we will use this convention 
in the following discussion of the behavior of the EOM self-energy. 
As typically seen in Fig.\ 9 of Ref.\ \cite{busser}, 
the results in Ref.\ \cite{chiappe} have 
also been affected by the singularity, 
although the position of the pole might shift 
due to the absence of the rotation symmetry of the spin 
in the convention used in Ref.\ \cite{chiappe}.

The presence of the $1/\omega$ singularity 
of the EOM self-energy for the cluster 
of an odd size can be deduced along the discussion 
similar to that in the section V of Ref.\ \cite{busser}.
Using the SU(2) symmetry of the axial charge 
holding in the electron-hole symmetric case
(see Appendix \ref{sec:axial_charge}),
the cluster Green's function can be expressed as, 
\begin{eqnarray}
&& 
\!\!\!\!\!\!\!\!\!
g_{{\rm cl};ij}(\omega) 
= 
\frac{S_g+1}{2S_g+1} 
\sum_{r'}\,
C_i(1/2,S_g+1/2;r'|0,S_g;r_g )
\nonumber \\ 
&&
\rule{2.6cm}{0cm}
\times\, C_j(1/2,S_g+1/2;r'|0,S_g;r_g ) 
\nonumber \\ 
&&
\rule{2.5cm}{0cm}
\times \left\{ 
{1 \over \omega - \omega_{S_g}^+(r')}  
\,+ \,{ (-1)^{i+j} \over \omega + \omega_{S_g}^+(r') }
\right\} 
\nonumber \\ 
&& \nonumber \\ 
&&
\rule{1cm}{0cm}
+\, 
\frac{S_g}{2S_g+1} 
\sum_{r'}\,
C_i(1/2,S_g-1/2;r'|0,S_g;r_g )
\nonumber \\ 
&&
\rule{2.9cm}{0cm}
\times\, C_j(1/2,S_g-1/2;r'|0,S_g;r_g ) 
\nonumber \\ 
&&
\rule{2.5cm}{0cm}
\times \left\{ 
{1 \over \omega - \omega_{S_g}^-(r')}  
\,+ \, { (-1)^{i+j} \over \omega + \omega_{S_g}^-(r') }
\right\} 
\;,
\nonumber \\ 
\label{eq:Lehmann}
&&
\\
&& 
\!\!\!\!\!\!\!
\omega_{S_g}^{\pm}(r') \ = \ E_{1/2,S_g \pm 1/2}(r') \, - \, E_{0,S_g}(r_g)
\label{eq:excitation}
\;.
\end{eqnarray}
Here $S_g$, $J_g$, and $r_g$ are the quantum numbers of the ground state, 
the energy of which is denoted by $E_{J_g,S_g}(r_g)$. 
Note that $J_g$ must be zero for $U>0$.
Furthermore, for $U>0$, 
the excitation energy must be finite $\omega_{S_g}^{\pm}(r')>0$ 
in the case of $N \geq 2$ as long 
as the cluster size $N_{\rm cl}$ is finite. 
Thus, for $N \geq 2$, 
the full Green's $\widehat{g}_{\rm cl}(\omega)$ does 
not have a pole at $\omega=0$ independent of 
whether $N_{\rm cl}$ is even or odd.
In contrast,
the noninteracting Green's 
function $\widehat{g}_{\rm cl}^0(\omega)$
does has a pole at $\omega=0$ for odd $N_{\rm cl}$, 
while it does not for even $N_{\rm cl}$.
Therefore the EOM self-energy $\widehat{\Sigma}_{\rm cl}(\omega)$, 
which can be calculated using Eq.\ (\ref{eq:DysonEOM}), 
must have a pole at $\omega=0$ for odd $N_{\rm cl}$ 
because the structure of the energy spectrum 
of $\widehat{g}_{\rm cl}(\omega)$ and 
that of $\widehat{g}_{\rm cl}^0(\omega)$ are 
completely different.
On the other hand,  for even $N_{\rm cl}$ the 
the energy spectrum of  
 $\widehat{g}_{\rm cl}(\omega)$ and 
 $\widehat{g}_{\rm cl}^0(\omega)$ belong to 
 the same category having no zero mode, 
 so that  $\widehat{\Sigma}_{\rm cl}(\omega)$ does not have  
 a singularity at $\omega=0$ and shows a behavior 
 consistent with that of 
 the Fermi liquid  Eq.\ (\ref{eq:Fermi_ground}). 
Therefore, 
the presence of the zero-energy pole can be summarized 
as Table \ref{tab:zero_mode} 
 (see also Appendix \ref{sec:EOM_N=1} for $N=1$). 
\begin{table}
\begin{ruledtabular}
\caption{ Pole at $\omega=0$ for the interacting sites of $N \geq 2$
\label{tab:zero_mode} 
}
\begin{tabular}{l|ccc} 
Cluster size \rule{0.4cm}{0cm} 
& $\widehat{g}_{\rm cl}(\omega)$  
& $\widehat{g}_{\rm cl}^0(\omega)$ 
& $\widehat{\Sigma}_{\rm cl}(\omega)$ \\
\hline
Even $N_{\rm cl}$  & None & None & None \\ 
Odd $\,N_{\rm cl}$ & None  & Yes & Yes \\ 
\end{tabular}
\end{ruledtabular}
\end{table}
Note that in the case of \lq\lq Yes", 
it does not necessary mean that all the matrix elements 
have the pole.

Using the EOM self-energy with the Dyson equation 
which connects the cluster to the remaining system, 
an approximate Green's function $G_{1N}^+$ can be calculated.
Alternatively, the same $G_{1N}^+$  can also be calculated  by
substituting the $N\times N$ partitioned part 
 of $\,\widehat{\Sigma}_{\rm cl}$,  which corresponds to 
 the interacting region, into the 
 Dyson equation of $N \times N$ form Eq.\ (\ref{eq:Dyson}),
 and then taking the inverse of the matrix.
Therefore, 
if one uses the cluster of odd $N_{\rm cl}$ for calculating 
 the value of $G_{1N}^+(\omega)$  at $\omega=0$,
one obtains the result \cite{busser}; 
$G_{1N}^+(0)=0$ and $g_{N}^{\phantom{\dagger}}=0$  for odd $N$ 
because of the zero-energy pole of $\,\widehat{\Sigma}_{\rm cl}(\omega)$. 
In contrast, if one uses the cluster of even $N_{\rm cl}$ in 
the same situation, the result must be different. 
This also suggests that the EOM is not suitable 
for studying the low-temperature properties.

\subsection{Interpretation of the zero-energy pole of 
$\,\widehat{\Sigma}_{\rm cl}$ }

The oscillatory behavior caused by the cluster size  $N_{\rm cl}$ 
has been seen also in the NRG calculation \cite{Wilson,KWW}. 
For even $N_{\rm cl}$, 
the ground-state is singlet and the self-energy can be expanded 
with respect to $\omega$ at the Fermi energy $\omega=0$, 
which agrees with the local Fermi-liquid theory 
of Nozi\`{e}res \cite{Nozieres}
and the perturbation theory of Yamada and Yosida \cite{YamadaYosida}. 
For odd $N_{\rm cl}$,
the ground state is doublet, and  
the self-energy has the zero-energy pole. 
Nevertheless, the energy spectrum of the low-lying states
shows the correct Fermi-liquid behavior 
even for odd $N_{\rm cl}$ \cite{KWW}.
Furthermore, the residue of the zero-energy pole 
decreases with increasing $N_{\rm cl}$ and  
vanishes in the  $N_{\rm cl} \to \infty$ limit.
Therefore, it is quite dangerous to examine 
the low-temperature properties only 
by the Green's function obtained with 
a finite cluster of an odd number sites $N_{\rm cl}$.

Similar oscillation takes place in an antiferromagnetic Heisenberg chain.
The spin susceptibility of the chain of an odd number sites has a
 Curie term. 
Since the weight of the Curie term is proportional to
$1/N_{\rm cl}$ and vanishes at $N_{\rm cl} \to \infty$, 
usually the cluster of an even number sites is used for studying
the properties in the thermodynamic limit,  
although the cluster of an odd number sites can also be used 
if the contribution of the Curie term is subtracted properly.

\section{Conclusion}
\label{sec:summary}

In conclusion, 
the EOM self-energy obtained for a finite cluster 
of an odd number of sites has the zero-energy pole, 
the residue of which must vanish in the limit of 
large cluster $N_{\rm cl}\to \infty$.   
The absence of the unitarity-limit transport 
reported in Refs.\ \cite{chiappe} and \cite{busser}
can be explained to be caused by a naive use of 
this singular self-energy.

\appendix

\section{Axial charge}
\label{sec:axial_charge}

In the electron-hole symmetric case,
the Hamiltonian $\widehat H$ commute with 
the total axial charge defined by \cite{CoxZawadowski}
\begin{eqnarray}
\widehat J^2 &=&  \widehat J_z^2 
+ ( \widehat J_+ \widehat J_- + \widehat J_- \widehat J_+)/2 \;, \\
\widehat J_z &=& \sum_{i} 
\,\frac{1}{2}\, 
\left(\,
c_{i\uparrow}^{\dagger}
c_{i\uparrow}^{\phantom{\dagger}}
+
c_{i\downarrow}^{\dagger}
c_{i\downarrow}^{\phantom{\dagger}} -1
\,\right) \;, \\
\widehat J_+ &=& \sum_{i} (-1)^{i}\,
c_{i\uparrow}^{\dagger}
c_{i\downarrow}^{\dagger} \;, \\
\widehat J_- &=& \sum_{i} (-1)^{i}\,
c_{i\downarrow}^{\phantom{\dagger}}
c_{i\uparrow}^{\phantom{\dagger}} \;.
\end{eqnarray}
Note that the $z$ component corresponds to the charge of 
the U(1) symmetry as $\widehat Q = 2 \widehat J_z$.
The operators  $\widehat J_z$ and $\widehat J_{\pm}$  satisfy 
the commutation relations of the angular momentum 
as the total spin operators $\widehat S_z$ and $\widehat S_{\pm}$ do.
Thus, the eigenstates can be classified according to 
the quantum numbers corresponding to 
 $\widehat J^2$, $\widehat J_z$
 $\widehat S^2$, and $\widehat S_z$;
\begin{equation}
\widehat H\, |J,J_z,S,S_z\, ; r  \rangle 
\, = \, E_{J,S}(r)\, |J,J_z,S,S_z\, ; r  \rangle \;.
\end{equation}
Eq.\ (\ref{eq:Lehmann}) can be derived using the Wigner-Eckart theorem 
for the spin and for the axial charge, 
\begin{eqnarray}
&&
\!\!\!\!\!\!\!\!\!\!\!\!\!\!\!\!\!\!
 \langle 
 J,J_z,S,S_z\, ; r  
 | c_{i\sigma}^{\dagger} |
  J',J_z',S',S_z'\, ; r' 
 \rangle 
 \nonumber \\
&=& 
 \langle S', S_z' ; 1/2, \sigma | S, S_z \rangle \, 
 \langle J', J_z' ; 1/2, 1/2 | J, J_z \rangle
 \nonumber \\
&& \times \, C_i(J,S;r|J',S';r' )  \;,
\end{eqnarray}
where the Clebsh-Gordan coefficient appears for the total spin
 $\langle S', S_z' ; 1/2, \sigma | S, S_z \rangle$ 
and for the total axial charge $\langle J', J_z' ; 1/2, 1/2 | J, J_z \rangle$.
Note that the invariant matrix element $C_i(\alpha'|\alpha)$ has 
the following properties against the exchange of the arguments 
$\alpha'$ and $\alpha$, 
\begin{eqnarray}
&& 
\!\!\!\!
C_i(J,S-\frac{1}{2};r'|J-\frac{1}{2},S;r ) 
\nonumber \\
&& 
\!\!\!\!
= (-1)^{i}
 \sqrt{2S+1 \over 2S}
\sqrt{2J \over 2J+1} \,
C_i(J-\frac{1}{2},S;r|J,S-\frac{1}{2};r') \,,
\nonumber \\
\\
&& 
\!\!\!\!
C_i(J,S+\frac{1}{2};r'|J-\frac{1}{2},S;r ) 
\nonumber \\
&& 
\!\!\!\!
= (-1)^{i+1}
 \sqrt{2S+1 \over 2S+2}
 \sqrt{2J \over 2J+1} \, 
C_i(J-\frac{1}{2},S;r|J,S+\frac{1}{2};r') \,.
\nonumber \\
\end{eqnarray}

\section{EOM results for $N=1$}
\label{sec:EOM_N=1}

A cluster consisting of 
a single Anderson impurity at the center $i=0$,
the left lead of $N_L'$ sites at $i<0$, and 
the right lead of $N_R'$ sites at $i>0$ can be separated 
into two independent parts by a Unitary transformation 
$a_{i\sigma} = (c_{i\sigma}+c_{-i\sigma})/\sqrt{2}$ 
and  
$b_{i\sigma} = (c_{i\sigma}-c_{-i\sigma})/\sqrt{2}$ 
   for $i=1,2,3, \ldots$ 
when $N_L'=N_R'$ ($\equiv N_b'$). 
Only the orbits of $a_{i\sigma}$ couple with 
 the Anderson impurity to form a cluster 
of the size $N_a'$ ($\equiv N_b' + 1$),
and the orbits of $b_{i\sigma}$ form a noninteracting band 
of the size $N_b'$. 
Thus, the EOM self-energy $\Sigma_{\rm cl}(\omega)$ can be 
calculated using only the cluster of the interacting part, 
and the properties corresponding to Table \ref{tab:zero_mode} 
can be shown to hold by replacing $N_{\rm cl}$ with $N_a'$: 
   $\Sigma_{\rm cl}(\omega)$ is zero at $\omega=0$ and 
   conductance reaches the unitarity limit for even $N_{a}'$, 
   while $\Sigma_{\rm cl}(\omega)$ has a pole at $\omega=0$ and 
   the conductance becomes zero for odd $N_{a}'$.
Obviously, the conductance calculated from  
the self-energy having the zero-energy pole contradict with 
the correct unitarity-limit behavior known 
in the thermodynamic limit $N_{a}' \to \infty$. 
Note that in Ref.\ \cite{chiappe},
the self-energy for even  $N_{a}'$ is used for $N=1$,
while the self-energy having a pole around Fermi energy 
is used for $N=3,\,5,\,7$ (see also Fig.\ 9 of Ref.\ \cite{busser}).

\end{document}